\documentstyle[11pt]{article}

\def\suq{su$_{q}$(2)}

\newcommand{\sect}[1]{\setcounter{equation}{0}\section{#1}}
\newcommand{\subsect}[1]{\subsection{#1}}

\def\be{\begin{equation}}
\def\ee{\end{equation}}
\def\bea{\begin{eqnarray}}
\def\eea{\end{eqnarray}}

\def\v#1{|{#1}\rangle}
\def\vc#1{\langle{#1}|}
\def\vu#1{|\underline{#1}\rangle}

\def\1{\'{\i}}
\def\R{{\rm I\kern-.2em R}}

\def\jp{J_+}
\def\jm{J_-}
\def\jpm{J_\pm}

\def\j{J_z}

\begin{document}

\thispagestyle{empty}
 \

\
\vspace{3cm}

\begin{center}
{\LARGE{\bf{ On the spectrum of a Hamiltonian}}}

{\LARGE{\bf{ defined on $su_q(2)$}}}

{\LARGE{\bf{and quantum optical models}}}

\end{center}

\bigskip\bigskip

\begin{center} Angel Ballesteros$^\dagger$ and Sergei M.
Chumakov$^\ddagger$
\end{center}

\begin{center} {\it {  $^\dagger$ Departamento de F\1sica, Universidad
de Burgos} \\   Pza. Misael Ba\~nuelos,
E-09001-Burgos, Spain}
\end{center}

\begin{center} {\it {  $^\ddagger$  Departamento de F\1sica, Universidad de
Guadalajara\\ Corregidora 500, 44420,  Guadalajara, Jal., M\'exico}}
\end{center}

\bigskip
\bigskip
\bigskip

\begin{abstract}

Analytical expressions are given for the eigenvalues and
eigenvectors of a Hamiltonian with \suq\  dynamical symmetry.
The relevance of such an operator in Quantum Optics is discussed.
As an application, the ground state energy in the Dicke model is
studied through \suq\ perturbation theory.

\end{abstract}

\bigskip
\bigskip
\bigskip
\bigskip
\bigskip
\bigskip



\newpage

\sect{Introduction}

Many models in Quantum Optics, such as Raman and Brillouin scattering,
parametric conversion and the interaction of two-level atoms with a
single-mode
radiation field (Dicke model), can be described by interaction Hamiltonians
of  the
form (see, {\it e.g.} [1-4])
\be
H=   \left( \begin{array}{ccccc}
       0    & A_l    & 0        & \ldots  &  0      \\
       A_l  &  0     & A_{l-1}  & \ldots  &  0      \\
     \ldots & \ldots & \ldots   &  \ldots & \ldots  \\
       0    & \ldots & A_{-l+2} &   0     & A_{-l+1}  \\
       0    & \ldots &   0      &  A_{-l+1} &  0
    \end{array}\right).
\label{int}
\ee
The dimension $d=(2l+1)$ of this matrix is by no means small (for instance,
in  the Dicke model, $2\,l$ is just the number of atoms considered).
Therefore, the finding of analytical expressions for the corresponding
eigenvalues and eigenvectors of $H$ is essential in order to solve the
dynamics of the model. It is also important to point out that, in some
cases,
$H$ can be seen as a perturbation of the $J_x$ generator of an underlying
$su(2)$ dynamical symmetry. This fact has been successfully used in order
to describe many features of these models \cite{Koz}, and it will be also
relevant in what follows.

In this paper we show that the \suq \  quantum algebra (see {\it e.g.}
[5-13])
can be used to
define a Hamiltonian of the type (\ref{int}) in a natural way. Such
Hamiltonian
is introduced as a simple function of the \suq\ generators having as
non-deformed
limit the $J_x$ generator of the $su(2)$ algebra. By considering the
well-known
representation theory of \suq (which is revisited in Section 2), the
Hamiltonian is defined and its eigenvalues and eigenvectors are found
(Section 3). The spectrum obtained is essentially anharmonic; thus we have
a new exactly solvable nonlinear quantum model with
\suq\  dynamical symmetry.

In Section 4,  relevant
Clebsch-Gordan coefficients for both the bare and the dressed basis are
considered. Finally, in Section 5 we present an application of the previous
analytical results to finding the energy of the ground state of
the Dicke model by making use of a perturbation theory around the \suq\
Hamiltonian. This preliminary study shows that the
$q$-deformed
$J_x$ operator here introduced is physically meaningful in the context of
the  quantum optical Hamiltonians mentioned above (compare with the results
given in {\it e.g.} [2-4]), and the explicit dynamical features of the
\suq\  Dicke model will be fully developed in a forthcoming paper.
In this work, we shall provide the basic algebraic properties that are
needed in order to solve it explicitly. It is interesting to stress that
some of these properties show new (to our knowledge) features of the \suq\
algebra, all of them related to the deformed $J_x=(J_+ + J_-)/2$
generator.


\sect{The \suq\ algebra}

Let the operators $J_z\,,J_{\pm}$
generate the quantum  algebra \suq\  with the coproduct \cite{KR}-\cite{Rosso}
\be
\Delta(J_{\pm}) = J_{\pm}\otimes q^{-J_z} + q^{J_z}\otimes J_{\pm}\,,
\label{copm}
\ee
\be
\Delta(J_{z}) = J_{z}\otimes 1 + 1\otimes J_{z}\,.
\label{coj}
\ee
Deformed commutation rules consistent with the previous map are given in
the form:
\be
[\j,\jpm]=\pm\,\jpm,\qquad [\jp,\jm]=[2\,\j],
\label{conm}
\ee
where $[x]:=\frac{q^x - q^{-x}}{q - q^{-1}}$ and $q=e^{z/2}$. We shall
assume that $q$ is not a root of unity, and we shall recover ``classical"
results  when
$q\to 1$.

Let us introduce, firstly, the ``bare'' basis of  eigenvectors of $J_z$,
\be
2J_z \v{l,m} = 2m\v{l,m}\,.
\label{jm}
\ee
In this basis, the $(2\,l+1)$-dimensional irreducible representation
of \suq\ is given by (\ref{jm}) and
\be
J_\pm\v{l,m} =
\sqrt{[l\mp m][l\pm m+1]}\v{l,m+1}.
\label{jmpm}
\ee
Throughout the paper we will use the following (standard) notation for
$q$-numbers,
\be
   [n] = q^{-n+1} + q^{-n+3} + q^{-n+5} \ldots + q^{n-1}
    = \frac{q^n-q^{-n}}{q-q^{-1}}\,,
   \quad [0] =0, \quad [-n] = -[n].
\label{q}
\ee
We shall also introduce the symbols
\be
   [n,m] := q^{-(n-1)m} + q^{-(n-3)m} + q^{-(n-5)m} \ldots + q^{(n-1)m}
         = [nm]/[m]\,.
\label{newq}
\ee
Thus, $\lim_{q\rightarrow 1}[n,k]=n$, and
\be
   [n,1]=[n]\,, \quad [k,0]=k\,,
   \quad [2,k] =q^k + q^{-k} = [2k]/[k]\, .
\label{new}
\ee


\sect{The Hamiltonian}

Let us consider the following
\suq\ operator
as a Hamiltonian:
\be
     H= q^{J_z/2} \,(J_+ + J_-) \,q^{J_z/2}\,.
\label{h}
\ee
We stress that the $q\to 1$ limit of $H$ is just $2\,J_x$, but $H$ is {\it
not}
the standard way to introduce the deformed $J_x$ operator in \suq.
On the other hand, one can easily check that, from the definition (\ref{h})
and the maps (\ref{copm})-(\ref{coj}), the coproduct of $H$
can be deduced (note that $\Delta$ is an algebra homomorphism:
$\Delta(X\,Y)=\Delta(X)\,\Delta(Y)$):
\be
     \Delta(H) = H\otimes 1  +  q^{2J_z}\otimes H\,.
\label{coh}
\ee
Such a form of the coproduct is known (see, {\it e.g.},
\cite{Rosso}). In the $(2\,l+1)$-dimensional representation $D_l$,
$H$ thus takes the form (\ref{int})  with
\be
   A_m = q^{m-1/2}\sqrt{[l\!+\!m][l\!-\!m\!+\!1]}\,.
\label{me}
\ee
In particular, when $l=1$ we have,
\be
  D_1(H)=  \left( \begin{array}{ccc}
       0                 &  q^{1/2}\sqrt{[2]}  & 0                   \\
       q^{1/2}\sqrt{[2]}  &  0                 & q^{-1/2}\sqrt{[2]}   \\
       0                 & q^{-1/2}\sqrt{[2]}  & 0
    \end{array}\right).
\label{h1}
\ee
A straightforward computation shows that the spectrum of this operator is
$[2],\,0,\,-[2]$. The corresponding normalized eigenvectors are
\be
  \vu{1,\pm1} = \frac{1}{\sqrt{2[2]}}\left( \begin{array}{c}
        q^{1/2} \\  \pm \sqrt{[2]} \\  q^{-1/2}
     \end{array}\right)\,, \qquad
  \vu{1,0} = \frac{1}{[2]}\left( \begin{array}{c}
        q^{-1/2}\sqrt{[2]} \\ 0 \\ - q^{1/2} {\sqrt{[2]}}
     \end{array}\right)      \,.
\label{eig1}
\ee
Due to the richness of the structures underlying quantum deformations, this
remarkable anharmonic deformation of the $J_x$ Hamiltonian can be explicitly
solved for arbitrary
$l$ as follows.


\subsect{Spectrum and eigenvectors}

 It can be proven (by
induction and using the coproduct to construct higher dimensional
representations from the three-dimensional one considered before) that the
spectrum of this operator for a given $l$ is just $[2m]$, with
$m=-l,\dots,l$.

Moreover, the  eigenvector $\vu{l,l}'$ corresponding to the highest
eigenvalue
$[2\,l]$ is given by the following components in the bare basis $\v{l,m}$:
\be
   \alpha_{ml} \equiv  \langle l,m\vu{l,l}'
     = q^{m(l-1/2)} \sqrt{\frac{[2l]!}{[l\!+\!m]![l\!-\!m]!}}\,.
\label{EV0}
\ee
Here, the prime means that a special normalization is accepted,
temporarily, where $\alpha_{ll}=q^{l(l-1/2)}=1/\alpha_{-l-l}$.
The next eigenvector $\vu{l,l-1}'$ with the eigenvalue $[2l\!-\!2]$ is
given  by
\be
  \alpha_{m,l-1} =
   \langle l,m\vu{l,l-1}' = \alpha_{ml}\, q^{-2m+1}\sqrt{[2l]}\,
   \left( 1 - q^{l+m-1} \,[2]\, \frac{[l-m]}{[2l]} \right)  \,.
\label{ev1}
\ee
Finally, an arbitrary eigenvector with eigenvalue $\,[2m]$ can be deduced,
namely
\bea
 && \!\!\!\!\!\!\!\!\!\!\!\!\!\!\!\!\!
\alpha_{mn} = \langle l,m\vu{l,n}' = \alpha_{ml}\,
       q^{(l-n)(l-n-2m)}\,
        \sqrt{\frac{[2l]!}{[l\!-\!n]![l\!+\!n]!}}\, \\
  && \!\!\!\!\!\!\!\!\!\!\!\!\!\!\!\!\!\!\!\!\!\!\!\!\!\!\!\!
\times
     \sum_{j=0}^{l-n}  (-1)^j q^{-j(l-m-2n) + j(j+1)/2}
       \,\frac{[2(l\!-\!n)]!!}{[j]![2(l\!-\!n\!-\!j)]!!}
       \,\frac{[l\!-\!m][l\!-\!m\!-\!1]\ldots[l\!-\!m\!-\!j\!+\!1]}
                {[2l][2l\!-\!1]\ldots[2l\!-\!j\!+\!1]}\,,
\label{EV}
\eea
where $[2n]!!:=[2n]\cdot [2n -2]\cdots [2]$.


\subsect{Normalization}

These eigenvectors can be easily normalized in terms of $q$-numbers.
For the ground and top states we have,
\be
        \vu{l,l} ={\cal N}_{l,l}^{-1}\,\vu{l,l}'\,, \ \mbox{where} \
   {\cal N}_{l,l}^{2}  =  {\cal N}_{l,-l}^{2}=
   \,'\langle\underline{l,l}\vu{l,l}'
   =q^{-l(l-1/2)}\,\sum_{k=0}^{2l} \frac{q^{k(2l-1)}[2l]!}{[k]![2l\!-\!k]!}
      \nonumber
\label{norm0}
\ee

A simple calculation gives,
\be
\label{2^N}
  {\cal N}_{l,l}^{2}    = \prod_{k=0}^{2l-1} (q^k+q^{-k}) =
    2\,[2]\,(q^2+q^{-2})\cdot\ldots \cdot
    (q^{2l\!-\!1}\!+\!q^{-\!2l\!+\!1})= \prod_{k=0}^{2l-1}{[2,k]}  \,.
\label{ground}
\ee

The excited states can be normalized as follows:
\be
\label{2^2l}
   {\cal N}_{l,n}^{2}   = \,^{\prime}\langle\underline{l,n}
   \vu{l,n}^{\prime}\,  = \frac{1}{q^{2n} + q^{-2n}}\,
      \prod_{k=-l+n}^{l+n} \left( q^k + q^{-k}\right)
=\frac{1}{[2,\,2n]} \prod_{k=-l+n}^{l+n}{[2,k]}\,.
\label{exc}
\ee
For instance,
\be
{\cal N}^2_{1/2,1/2} = {\cal N}^2_{1/2,-1/2}= 2\,,\quad
   {\cal N}^2_{1,1}={\cal N}^2_{1,-1} = 2\,[2]\,, \quad
   {\cal N}^2_{1,0}= [2]^2\,,\quad \ldots
\label{exn}
\ee
Note that all of the coefficients $ {\cal N}^2_{l,n} $ go to  $2^{2l}$
in the limit $q\rightarrow 1$. Finally, if we denote the entries of the
normalized eigenvectors as
\be
     {\cal A}_{mn}^{l} = \alpha_{mn}^{l}/{\cal N}_{ln}\,,
\label{db2}
\ee
the following relations hold,
\be
    {\cal A}^{l}_{mn}(q) = (-1)^{l-n}{\cal A}^{l}_{-m,n}(1/q)\,, \qquad
      {\cal A}^{l}_{mn}(q) = (-1)^{l-m}{\cal A}^{l}_{m,-n}(q)\,.
\ee
They generalize the known symmetry of the non-deformed $su(2)$ case ($q=1$)
by involving the transformation $q\rightarrow q^{-1}$.


\subsect{Orthogonality relations}

For physical applications (like the calculation of mean values) is
desirable to
further develop the previous ``$q$-arithmetics''. With this goal in mind,
let  us
note that  the quantities $ |\alpha_{ml}|^2 $  from Eq.\ (\ref{EV0})
play the role of the binomial coefficients.
In particular, the moments of this deformed binomial distribution, {\it
i.e.}  the mean values,
\be
   <Y_k>_{2l} \; \equiv \sum_{k=0}^{2l}  |\alpha_{k-l,l}|^2\, Y_k\,.
\label{yk}
\ee
can be calculated.
With this definition, the equation (\ref{norm0}) is then rewritten in the
form,
\be
<1>_{2l}\; \equiv \, \prod_{s=0}^{2l-1} (q^s+q^{-s}\,)  \,.
\label{bin}
\ee
Moreover, one has the following deformed formulas $(0\leq j,M \leq 2l)$ for
the
binomial  distribution:
\bea
      & & \!\!\!\!\!\!\!\!\!\!\!\!\!\!\! <q^{-2\,j\,k}>_{2l} \;=\,
q^{-2\,j\,l}\,
        \prod_{s=-j}^{2l-j-1} (q^s + q^{-s})\,, \\
      & &
\!\!\!\!\!\!\!\!\!\!\!\!\!\!\!
<[k][k\!-\!1]\ldots[k\!-\!M\!+\!1]\,q^{-k(2j+M)}>_{2l}
\;= \,
        q^{-2Mj - M(M\!+\!1)/2}\, <q^{-2\,j\,k}>_{2l-M} \label{for1}\\
      & & \qquad =  [2l][2l\!-\!1]\ldots[2l\!-\!M\!+\!1]\;
          q^{-j(2l+M)-M(M+1)/2}\,\prod_{s=-j}^{2l-M-1-j}(q^s+q^{-s})\,.
\label{for2}
\eea
These formulas are equivalent to the fact that the dressed
vectors form an orthonormal basis.


\sect{Clebsch-Gordan coefficients for the dressed basis}

We can decompose the tensor product of the dressed vectors
(\ref{EV}),(\ref{db2})
into the irreducible parts:
\be
  \vu{l,m}\otimes\vu{1,i} = \sum_{j=l-1}^{j=l+1}
    \underline{C}^j_{l1,mi} \v{\overline{j,m+i}}_{12}\,,
\label{cgd}
\ee
where $ \underline{C}^j_{l1,mi}$ are the  Clebsch-Gordan (C.G.)
coefficients  for the dressed basis.
However, the vector $ \vu{l,m}\otimes\vu{1,i}$
in the left-hand side and vectors
  $\v{\overline{j,m+i}}_{12}$  in the right-hand side
are eigenvectors of $\underline{J}_x$  in the
corresponding representations only for $i=0$. Thus, only the coefficients
$ \underline{C}^j_{l1,m0}$ are of interest. For $l=1$, they are given
as
\bea
  \underline{C}^2_{11,10}   = \sqrt{\frac{q^3+q^{-3}}{[4]}},\quad
    & \underline{C}^1_{11,10}  = - \sqrt{\frac{1}{q^2+q^{-2}}},\quad
    & \underline{C}^0_{11,10}  =0, \\
  \underline{C}^2_{11,00}   =  \sqrt{\frac{[4]}{[2][3]}},\quad
    & \underline{C}^1_{11,00}  = 0 ,\quad
    & \underline{C}^0_{11,00}  = - \sqrt{\frac{1}{[3]}}, \\
  \underline{C}^2_{11,-10}  = \underline{C}^2_{11,10} ,\quad
    & \underline{C}^1_{11,-10} = -\underline{C}^1_{11,10} ,\quad
    & \underline{C}^0_{11,-10} = 0.
\eea
They thus differ from the coefficients in the bare basis \cite{Clebsch}
(though have the same limit $q\rightarrow 1$).

By rewriting the definition,
\be
\label{CGdr}
 \underline{C}^{j}_{l1;\,p,0} = \,_{12}\langle\underline{ j,p}
   \vu{l,p}_1 \vu{1,0}_2\, ,
\ee
in terms of components of the dressed vectors,
and afterwards replacing explicitly the components of $\vu{1,0}_2$, we get
the relation
\be
\label{CGdr1}
    \underline{C}^{j}_{l1;p0}
   = \frac{1}{\sqrt{[2]}} \,\sum_{k=-j}^{j} {\cal A}^j_{kp}
   \left(  q^{-1/2}\,  C^j_{l1;k-1,1} \, {\cal A}^l_{k-1,p}
   -q^{1/2} \,  C^j_{l1;k+1,-1} \, {\cal A}^l_{k+1,p}
   \right)\,.
\ee
Here, ${\cal A}^l_{mn}=0$ if $|m|>L$ and
\be
C^j_{l1;m_1,m_2}=
\,_{12}\langle { j,m_1+m_2}
   \v{l,m_1}_1 \v{1,m_2}_2\, ,
\label{cgbare}
\ee
are C.G. coefficients in the
bare basis.  Another useful formula for the Clebsch-Gordan coefficients in
the dressed basis can be obtained as follows.
Starting with the expansion of the tensor product,
\be
        \vu{l,p}_1 \vu{1,0}_2 = \sum_j
          \underline{C}^{j}_{l1;\,p0} \vu{j,p}_{12}\,,
\ee
we can rewrite the dressed vectors in the bare basis,
$\vu{j,p}_{12}=\sum_{m=-j}^{j}{\cal A}_{mp}^j \v{j,m}_{12}$,
and the vectors $\v{j,m}_{12}$ in terms of the tensor product,
using bare C.G. coefficients,
\be
    \v{j,m}_{12} = \sum_{m_2=-1}^1
       C^j_{l1;\, m-m_2,m_2} \v{l,m-m_2}_1\v{1,m_2}_2\,.
\ee
We have,
\be
   {\cal A}^l_{m_1p}{\cal A}^1_{m_20} = \sum_{j=l-1}^{l+1}
     \underline{C}^{j}_{l1;\,p0} {\cal A}^j_{m_1+m_2,p} C^j_{l1;\, m_1m_2}.
\ee
Multiplying by  $C^k_{l_1l_2;\,m_1,m_2}$, making a summation under the
condition $m_1+m_2=m$,
using the orthogonality of the Clebsch-Gordan coefficients,
\be
   \sum_{m_2} C^k_{l_1l_2;\,m-m_2,m_2} C^j_{l_1l_2;\,m-m_2,m_2}
    = \delta_{jk}\,,
\ee
and, finally, replacing the explicit form of ${\cal A}^1_{m0}$,
we have the connection between the coefficients in the bare and dressed
basis,
\bea
\label{CGdr2}
  \underline{C}^{j}_{l1;\,p0}{\cal A}^{j}_{mp} &=&
   \frac{1}{\sqrt{[2]}} \,\left\{
    - {\cal A}^l_{m+1,p}\,q^{1/2} \,C^j_{l1;\,m+1,-1}
    + {\cal A}^l_{m-1,p}\,q^{-1/2}\, C^j_{l1;\,m-1,1}
    \right\}\,, \\
    & & \quad\quad  -j \leq m\leq j, \quad -l\leq p\leq l\,. \nonumber
\eea
For instance, using these formulas for $m=j$, we can express
the``dressed'' coefficients in terms of ``bare'' ones.

Finding components of the first three rows of the matrix ${\cal A}^l_{mn}$,
\bea
    {\cal A}^l_{ln} &=& \frac{q^{n^2-l/2}}{{\cal N}_{ln}}
       \sqrt{\frac{[2l]!}{[l-n]![l+n]!}}\,, \nonumber\\
    {\cal A}^l_{l-1,n} &=& \frac{q^{n^2-(3l-1)/2}}{{\cal N}_{ln}}
\sqrt{\frac{[2l]!}{[l-n]![l+n]!}}\,\frac{[2n]}{\sqrt{[2l]}}\,,\label{final}
\\
    {\cal A}^l_{l-2,n} &=& \frac{q^{n^2-l/2+1}}{{\cal N}_{ln}}
       \sqrt{\frac{[2l]!}{[l-n]![l+n]!}}\,
       \frac{q^{1-2l}[2n]^2-[2l]}{\sqrt{[2l][2l][2l-1]}}\,\nonumber,
\eea
and, from (\ref{CGdr2}) we arrive at the explicit expressions,
\bea
\underline{C}^{l+1}_{l1;p0}&=&\sqrt{\frac{[l+1+p][l+1-p][2,l+1+p][2,l+1-p]}
  {[2][2l+2][2l+1]}}\,,\nonumber\\
\underline{C}^{l}_{l1;p0}&=&-\frac{[2p]}{\sqrt{[2l+2][2l]}}\,\label{final2},
\\
  \underline{C}^{l-1}_{l1;p0}&=&-\sqrt{\frac{[l+p][l-p][2,l+p][2,l-p]}{
           [2][2l][2l+1]}}\,.\nonumber
\eea


\sect{Application to the Dicke model}

In order to demonstrate how our approach works we will apply it to the
Dicke model, which describes the interaction of a system of $N=2l$
two-level atoms with the quantum radiation field in an ideal cavity.
   (This model is mathematically equivalent to the three-photon
Hamiltonian describing three-wave mixing.)
   The Hamiltonian can be written in the matrix form (\ref{int}) with
the matrix elements
\be
    B_m = \sqrt{(l+m)(l+1-m)(s+1-m)},
\ee
where $s\geq l$  is a parameter ($2s$ is the excitation number which
is conserved in this model \cite{Koz}).
   We restrict ourselves with the case of the highest nonlinearity,  $s=l$.
(In the language of the three-wave mixing processes this corresponds to the
second harmonics generation.)
   We are specially interested in the limit of large $l$, which corresponds
to high photon numbers.

We will take the Hamiltonian
\be
   H_0=\Omega\,H,
\ee
with $H$ (\ref{h}) belonging to $su_q(2)$ as a zeroth-order Hamiltonian and
we will find the Dicke spectrum by using the perturbation theory. In the
present work we restrict ourselves with the energy of the ground state, for
the sake of simplicity. The values of $q$ and $\Omega$ are to be chosen. The
simplest way to fit them is to provide the coincidence of the points where
the matrix elements $B_m$ of the three-wave Hamiltonian and the matrix
elements
\be
  A_m =\Omega\, q^{m-1/2}\sqrt{[l-m][l+m+1]} =
 \Omega \,\vc{l,m+1} 2J_x \v{l,m}
\ee
of the Hamiltonian $H_0$ take their maximum values. It gives (for $s=l$)
\be
  \alpha= N\log q = \frac{3}{2} \log \frac{\sqrt{5}-1}{2} \approx -0.7218\,,
\ee
and the maxima of $B(m)$ and $A(m)$ occur in the point $m_0=-(l-1)/3$.
We chose the coefficient $\Omega$ to make equal the values
of $A_m$ and $B_m$ in their maxima. This gives
\be
    \Omega =\frac{4(N+1)^{3/2}}{\sqrt{27} [N+1]}\,.
\ee

At the next step, we find the approximation for the three-wave
Hamiltonian in the form:
\begin{equation}
\label{Hap}
    B_m \approx \Omega\, A_m\, \phi(m),\quad
    \phi (m) = 1 +\phi_1 \Delta - \phi_2 \Delta^2 + \phi^3 \Delta^3\,,
    \quad  \Delta = m-m_0\,.
\end{equation}
We thus restrict the expansion up to the third-order polynomial
$\phi(m)$. We can find explicitly the coefficients $\phi_j$ by comparing the
Taylor expansions for the matrix elements $B(m)$ and $A(m)$ around the
point $m_0$. We have:
\be
   B(m) =  \frac{2(N+1)^{3/2}}{\sqrt{27}} \,
   \left[  1 - \frac{27}{8}\,\left( \frac{\Delta}{N+1} \right)^2
             + \frac{27}{8}\,\left( \frac{\Delta}{N+1} \right)^3 +
             O\left( N^{-4} \right)
   \right],
\ee
and
\be
   A(m) =  \frac{[N+1]}{2} \,
   \left[1 - \frac{2\alpha^2}{\tanh^2\alpha}\,
              \left(\frac{\Delta}{N+1} \right)^2
      - \frac{4\alpha^3}{\tanh^2\alpha}\,
        \,\left( \frac{\Delta}{N+1} \right)^3 +
             O\left( N^{-4} \right)
   \right],
\ee
which determines the polynomial $\phi(\Delta)$:
\be
    \phi(\Delta) = 1
    - \left(\frac{27}{8} - \frac{2\alpha^2}{\tanh^2\alpha} \right)\,
         \left(\frac{\Delta}{N+1} \right)^2
    + \left(\frac{27}{8} + \frac{4\alpha^3}{\tanh^2\alpha} \right)\,
         \left(\frac{\Delta}{N+1} \right)^3 + O\left( N^{-4} \right)\,.
\ee
Now we may substitute $\Delta = m-m_0=J_z +(l-1)/3$ and rewrite
(\ref{Hap}) in the matrix form:
\be
     H\approx \Omega \left[ J_+ \phi(J_z-m_0) +  \phi(J_z-m_0) J_-  \right]
     = 2\Omega \left\{J_x , f(J_z)  \right\} .
\ee
Here $J_{\pm,z}$ are generators of $su_q(2)$ and $\{A,B\} =AB+BA$.
The new function $f(J_z)$ is also a polynomial of degree three, whose
coefficients can be easily found.
Now the ground state energy is approximately given as
\be
   \vc{\underline{-l,l}} H \v{\underline{-l,l}}
   \approx - \Omega [2l] \, \sum_{k=0}^3 f_k
           \vc{\underline{-l,l}}(J_z)^k\v{\underline{-l,l}}.
\label{gse}
\ee
Therefore, we have reduced the problem to the calculation of averages of
the powers of the operators $J_z$ (the moments) in the eigenstates of the
operator
$J_x$.

Though this problem can be solved for arbitrary eigenstates, here we
consider the ground states only. By using the results of the previous
sections we can write the generating function for these moments.
For arbitrary $p=\exp\mu$ and $q=\exp t$ we find:
\begin{equation}
\label{pq}
     \vc{\underline{-l,l}} p^{2J_z}\v{\underline{-l,l}} = \prod_{k=0}^{N-1}
     \frac{\cosh(\mu+k t)}{\cosh(k t)}.
\end{equation}
Now, let us introduce the notations
\begin{equation}
\label{Sk}
    S_k=\sum_{j=1}^{N-1} \tanh^k jt \approx
     \int_{1}^{N-1} dx \tanh^k xt +\frac{1}{2}\left[
       \tanh^k t + \tanh^k (N-1)t  \right]\,,
\end{equation}
where  we have used the Euler-Maclaurin summation formula.
Differentiating (\ref{pq}) with respect to $\mu$ we find,
\bea
    \vc{\underline{-l,l}}{2J_z}\v{\underline{-l,l}}      &=&  S_1\,, \\
    \vc{\underline{-l,l}}{(2J_z)^2}\v{\underline{-l,l}}  &=&
        S_1^2 -S_2 + N \,, \label{S}\\
    \vc{\underline{-l,l}}{(2J_z)^3}\v{\underline{-l,l}}  &=&
        S_1^3 - 3 S_2 S_1 + (3N-2) S_1 + 2 S_3 \,.
\eea
On the other hand, for the first three sums we have from (\ref{Sk}),
\bea
    S_1 &\approx & \frac{N}{\alpha} \,\log\cosh(\alpha-t)
       + \frac{\tanh(\alpha-t)}{2}\,,  \qquad t=\frac{\alpha}{N}=\log q, \\
    S_2 &\approx & N-1 - N\,\frac{\tanh(\alpha-t)}{\alpha}
        + \frac{\tanh^2(\alpha-t)}{2}\,, \\
    S_3 &\approx & \frac{N}{\alpha} \,\log\cosh(\alpha-t)
         - N\,\frac{\tanh^2(\alpha-t)}{2\alpha}
         + \frac{\tanh^3(\alpha-t)}{2}\,.
\eea

The combination of all these formulas gives the approximation for
the energy of the ground state.
Comparing with the numerical results, we can say that the accuracy for
the energy of the ground state is 1.5\% for 100 atoms ($N=100$)
and 0.35\% for 400 atoms. Note that it is the Maclaurin summation
formula (\ref{Sk}) that reduces the accuracy, which would be otherwise
much higher. However, it gives the correct asymptotic behaviour
when $N\rightarrow\infty$, which is sufficient for our goals.
   We may mention also that our method produces much better
accuracy than the analogous perturbation theory with common
$su(2)$ as a dynamical symmetry algebra \cite{Koz,Klimov}
or than the variational method with the $su(2)$  coherent states
as probe states \cite{Kar}.
The complete description of the spectrum and the dynamical analysis
of the Dicke model by means of the present approach will be given
elsewhere.

\bigskip
\medskip

\noindent
{\Large{{\bf Acknowledgments}}}

\noindent This work was partially supported by CONACYT, Mexico
(Project No. 465100-5-3927PE) and  by 
Junta de Castilla y Le\'on (Project CO2/399).
S.Ch. is grateful to the University of Burgos, Espa\~na for hospitality.
We thank V.P.\ Karassiov, A.B.\ Klimov and K.B.\ Wolf for interesting
discussions.



\end{document}